\documentclass{aa}
\usepackage{graphics}

\begin{document}

\thesaurus{22        
          (03.13.2;  
           03.19.2;  
           03.20.1;  
           04.03.1;  
           05.01.1;  
           08.02.3)} 

\title{The Hipparcos Transit Data: What, why and how?
\thanks{Based on observations made with the ESA Hipparcos
astrometry satellite}}

\subtitle{}

\author{C.F.~Quist
   \and L.~Lindegren}

\offprints{C.F.~Quist}

\institute{Lund Observatory, Box 43, SE-22100 Lund, Sweden
$<$fredrikq@astro.lu.se, lennart@astro.lu.se$>$}

\date{Received -- -- --; accepted -- -- --}

\titlerunning{Hipparcos Transit Data: What, why and how?}
\authorrunning{C.F.\ Quist \& L.\ Lindegren}

\maketitle

\begin{abstract}
The Hipparcos Transit Data are a collection of partially reduced, fully
calibrated observations of (mostly) double and multiple stars obtained
with the ESA Hipparcos astrometry satellite.  The data are publicly
available, as part of the CD-ROM set distributed with the Hipparcos and
Tycho Catalogues (ESA SP--1200, \cite{hip}), for about a third of the
Hipparcos Catalogue entries including all confirmed or suspected
non-single stars.  The Transit Data consist of signal
modulation parameters derived from the individual transits of the
targets across the Hipparcos focal grid. 
The Transit Data permit re-reduction of the satellite data
for individual objects, using arbitrarily complex object models in which
time-variable photometric as well as geometric characteristics may be
taken into account.  We describe the structure and contents of the
Transit Data files and give examples of how the data can be used.  Some
of the applications use standard astronomical software:  Difmap or AIPS
for aperture synthesis imaging, and GaussFit for detailed model
fitting.  Fortran code converting the data into formats suitable for
these application programs has been made public in order to encourage
and facilitate the use of Hipparcos Transit Data.
\end{abstract}

\section{Introduction}

In the course of its three-year lifetime the European Space Agency's
Hipparcos satellite performed some 13~million scans across the
$\sim\!118~000$ stellar objects on its pre-defined observing list.
For the vast majority
of targets the subsequent data reductions succeeded in determining,
accurately and without ambiguity, a small set of astrometric and
photometric parameters which contain practically all the useful
information on the objects.  In the simplest case of a non-variable
single star these data would be the position of the star at a certain
epoch, the trigonometric parallax, the two proper motion components, a
mean magnitude, and various statistics such as the mean errors of all
the parameters.  For thousands of variable, double and multiple stars,
the necessary additional parameters were determined without too much
problem.  All of these results are readily available in the Hipparcos
and Tycho Catalogues (ESA \cite{hip}).  For most objects and for most
applications of the results, there is no need to probe deeper into
the intricacies of the Hipparcos data acquisition and processing.

It was inevitable, however, that the observations of some (very small)
fraction of the objects could not easily be interpreted in terms of
standard models, and that inadequate, or even erroneous, object models
were applied in some cases.  Potentially very valuable information could
have been lost for these objects 
if only summary results, based on inadequate models, were
published.  This situation has been avoided by including a good
selection of {\it intermediate\/} results in the published catalogue.
Being only partially reduced these results are less dependent on
specific model assumptions, but still relatively simple to use thanks to
a careful calibration into the photometric and astrometric systems of
the final catalogue.  The published version of the Hipparcos and Tycho 
Catalogues includes six CD-ROMs, five of which contain mainly such
calibrated, partially reduced data.  The data fall in three distinct
categories:
\begin{itemize}
\item purely photometric information: the Epoch Photometry, consisting of
calibrated magnitudes for each transit of the targets across the field,
allowing a~posteriori analysis e.g.\ of stellar variability
(van Leeuwen et al.\ \cite{photom});
\item purely geometric information: the Intermediate Astrometric Data, 
which are the one-dimensional coordinates (`abscissae') of the targets
on reference great circles, each coordinate derived from several 
consecutive transits and referring to a (Hipparcos-specific) centroid
of the target (van Leeuwen \& Evans \cite{hipi});
\item combined photometric and geometric information:  the Transit Data
(TD), which are the Fourier coefficients (equivalent to amplitude and
phase) of the light modulation of the individual
transits for each of the targets across the $0\fdg 9\times 0\fdg 9$ main grid of
the Hipparcos instrument.
\end{itemize}
These mission products are fully described in Sects.~2.5, 2.8 and 2.9 
of Vol.~1 of the Hipparcos and Tycho Catalogues.
Examples using the Intermediate Astrometric Data have been
given by van Leeuwen \& Evans (\cite{hipi}).

While the TD thus provide information that is more detailed and on a
more fundamental level (in the sense of being closer to the satellite
`raw' data) than either the Epoch Photometry or the Intermediate
Astrometric Data, two important restrictions should be noted:  (1) TD
are only available for about a third of the targets, or some 38$\,$000
objects; (2) TD are based solely on
the results from one of the data reduction consortia, the Northern Data 
Analysis Consortium (NDAC; Lindegren et al.\ \cite{ndac}).  
By contrast, the Epoch Photometry and the Intermediate
Astrometric Data are available for virtually all the 118$\,$204
catalogue entries, and results from both NDAC and the Fundamental Astrometry 
by Space Techniques Consortium (FAST; Kovalevsky et al.\ \cite{fast}) 
were normally used.

Detailed satellite operations and data reductions are described
elsewhere (e.g., Perryman \& Hassan \cite{vol1}; 
Perryman et al.\ \cite{vol3}; Perryman et al.\ \cite{mission}; 
Kovalevsky et al.\ \cite{fast}; Lindegren et al.\ \cite{ndac};
Vols.~2--4 of ESA \cite{hip}; van Leeuwen \cite{floor}).  
It will however prove useful
to review how the TD were measured in order to understand their precise
meaning (and limitations), aiding in the full exploitation of the 
information.  Thus, Sect.~2 describes the convolution of the stellar
diffraction image with the modulation grid, and the resulting detector
signal modeled by the TD.  Section~3 is a detailed description of the
contents and format of the published TD files.  Two applications of the
TD using publicly available software are demonstrated, viz.\ aperture 
synthesis imaging (Sect.~4) and model fitting (Sect.~5).

\section{What are the Transit Data?}

\subsection{Availability of the data}
\label{sec:TD}

The TD are contained on a CD-ROM (Disk 6) in Vol.~17 of the Hipparcos and Tycho
Catalogues.  A formal description of the TD, including detailed
format specifications of the CD-ROM files, is found in Vol.~1, Sect.~2.9,
of the Hipparcos and Tycho Catalogues.  The TD come from an
intermediate step of the data reductions performed by NDAC.
In previous publications, the equivalent intermediate NDAC data
were referred to as `Case History Files' (S\"{o}derhjelm et al.\ \cite{ss92}).  
The data represent the scans (transits) of a selected number of targets
from the Hipparcos Catalogue (HIP), comprising 37$\,$368 systems
with 38$\,$535 different HIP entries.  (Some systems have two or three
HIP entries; cf.\ Fig.~\ref{fig:idx}.)  All Hipparcos stars that were
classified as double, multiple or suspected non-single are included in
the TD.  Also, stars having problematic solutions in the Hipparcos
Catalogue are included in the data set.  In order to provide reference
objects for calibrations or comparisons, several thousand ($\sim
5\,000$) `bona fide' single stars were also included in the TD.

\subsection{The modulated detector signal}
\label{sec:grid}

The physical layout of the Hipparcos optical instrument was a telescope with
two viewing directions in a plane perpendicular to the satellite's spin axis.
The two different fields of view were combined onto a single
focal surface by a special combining mirror.  The light from each
program star within either field of view was focused
onto the modulation grid, which was located in the focal surface.

The modulation grid consisted of a series of opaque and transparent
bands.  As the satellite spun, the diffraction image of each star traversed
the grid perpendicular to the bands, resulting in a periodic ($\simeq 7$~ms
period) modulation of the light intensity behind the grid.
The varying intensity was measured by the Image Dissector
Tube (IDT), a photomultiplier with an electronically steerable sensitive
spot (`Instantaneous Field of View', IFOV) of about 30~arcsec diameter.
During each `interlacing period' of $\simeq 130$~ms the IFOV would cycle
through the programme stars located within the
$0\fdg 9 \times 0\fdg 9$ field of view.  There were on average 4 to 5
programme stars within the field of view at any given time.

The telescope entrance pupil (for each of the two viewing directions) was
semi--circular, with diameter 0.29~m and with some central obscuration.
The Airy radius of the (ideal) diffraction image was thus around 0.5~arcsec
for an effective wavelength of 550~nm.  The modulation grid had a basic
period of 1.2074~arcsec, i.e.\ the separation between the centres of
adjacent transparent bands (slits).  The slit width was about 0.46~arcsec,
well matched to the Airy disk size and representing a compromise
between a sharp intensity maximum (requiring narrow slits) and
high photon throughput (requiring wide slits).

Since the grid was periodic, the detector signal, being the convolution of
the diffraction image with the grid transmittance, must also be a periodic
function of the image centroid coordinate on the grid.  Disregarding noise
and variations in detector sensitivity, etc., the signal therefore consisted
of a constant (DC) component plus modulated components having spatial
frequencies that are integer multiples of the fundamental grid frequency.
However, from the convolution theorem it follows that the signal cannot
contain higher spatial frequencies than were already in the diffraction image.
The maximum frequency in the diffraction image is given by the maximum 
separation of any two points in the pupil and the minimum detected
wavelength ($\simeq 350$~nm).  For a pupil diameter of 0.29~m this gives
a spatial period of 0.25~arcsec.  Thus the theoretically highest frequency
in the signal is four cycles per grid period (the `fourth harmonic').  The
fourth and third harmonics, with cycles of 0.3--0.4~arcsec, are however
very strongly damped by the slit width (0.46~arcsec), which causes an
averaging over more than one cycle for these components.  As a result,
the detector signal in practice contains only the first and second harmonics,
in addition to the DC (mean intensity) component.  Given the spatial frequency,
the detector signal is therefore completely parametrized by five numbers,
i.e.\ the DC term and the coefficients of the first two harmonics in the Fourier
series representing the periodic signal.  Furthermore, since Poisson (photon)
noise is by far the dominating noise source for most Hipparcos observations,
it can be shown that a proper estimate of these five Fourier coefficients
constitutes a {\it sufficient statistic\/} for the further estimation of the
photometric and geometric characteristics of the target, independent of its
complexity.
The TD contain precisely these Fourier coefficients along with the spatial
frequencies and other ancillary data.  From the viewpoint of statistical
estimation, practically no information was therefore lost by compressing
the raw photon counts (on average some 4500~bytes per transit and target)
into the five Fourier coefficients (included in the TD file).

The detector signal thus measured certain components of the
diffraction image moving over the modulation grid.  The diffraction
image, of course, had aberrations due to the imperfections of the telescope
and chromatic effects caused by the wavelength-dependent diffraction.  
Also, the sensitivity of the IDT varied across the field and over the
time.  All of these factors affected the
detector signal.  The user of the TD need not worry about this,
because the TD have been `rectified', which means all instrumental and colour
effects have been removed as far as possible by means of the various
calibrations produced in the data reductions.
Therefore, the TD should represent the response of an idealized, constant,
and well-defined instrument to the object.

\begin{figure}[t]
  \resizebox{\hsize}{!}{\includegraphics*{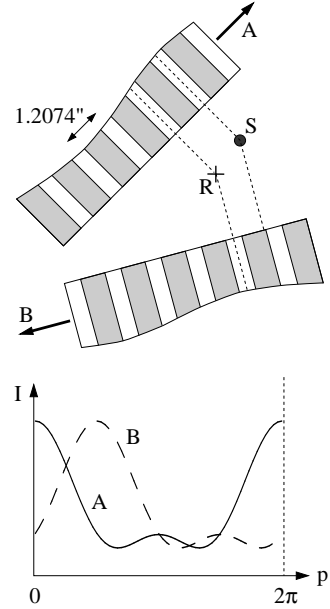}}
  \caption[ ]{This figure illustrates the definition of reference phase
    $p$ (as in Eq.~\ref{eq:ik}) by means of the reference point {\bf R}
    on the sky (with specified astrometric parameters).  $p=0$ at the
    instant when a slit is exactly centred on the reference point.
    In the lower diagram the curves A and B show the detector signal,
    as function of reference phase, produced by the same point source 
    {\bf S} when scanned in different directions.  The intensity maximum
    is displaced from $p=0$ depending on the separation of {\bf S} and 
    {\bf R} as projected on the grid, modulo the grid period 1.2074~arcsec.}
\label{fig:phase}
\end{figure}

\subsection{The reference point}
\label{sec:phase}

One of the key points of the TD is that the modulation phase in each scan
is expressed with respect to a well-defined reference point on the sky.
The astrometric parameters for the reference point (different for each
target) are also specified in the TD.  Usually the reference point
corresponds to the data given in the
Hipparcos Input Catalogue (HIC; Turon et al.\ \cite{hic}).
However, the phase calibration of the TD was made as if the coordinates
of the reference point were expressed in the Hipparcos
reference frame (nominally coinciding with the International Celestial
Reference System, ICRS; Feissel \& Mignard \cite{icrs}).  
The phase offset of the TD
from the reference point thus gives the differential correction to the position
from the reference point in the ICRS system.

The reference point in general has its own values for proper motion and
parallax, in addition to position.  These values are contained in the TD
`header record', see Sect.~\ref{sec:header}.  Of course, the real star also
has a position, proper motion, and parallax.  These two points move
independently on the sky, according to their respective proper motion
and parallax.  In a particular transit the phase is determined by 
the scan direction and the
distance between the two stars perpendicular to the slits.
Figure~\ref{fig:phase} illustrates how the phase varies in two different
scans across the star (S) and reference point (R).  In scan A the star
is in phase with the reference point, producing a light maximum
for $p=0$ (and more generally for $p=n2\pi$, where $n$ is an integer).
Scan B has the star out of phase with the reference point, resulting
in a light maximum about a third of a period later ($p \simeq 2$~rad).

The raw detector signal consists of a sequence of photon counts, $N_k$,
$k=1,2,\dots\,$, obtained in successive samples of $1/1200$~s integration
time.  The counts represent an underlying deterministic intensity
modulation which (after background subtraction and various calibrations)
is modeled as
\begin{equation}
   I_k = b_1 + b_2\cos{p_k} + b_3\sin{p_k} + b_4\cos{2p_k}
         + b_5\sin{2p_k}
  \label{eq:ik}  
\end{equation}
where $b_1$--$b_5$ are the Fourier coefficients given in the TD.
$I_k$ is the expected stellar count rate in sample $k$ (expressed in
photons per sample of $1/1200$~s) and $p_k$ is the reference phase of the
sample. The reference phase is defined by the following two conditions
(cf.\ Fig.~\ref{fig:phase}):  (1) that $p_k=0$ if a slit is exactly
centred on the reference point at the mid-time of sample $k$; and (2)
that $p_k$ is increasing from 0 to $2\pi$ over the time it takes the
grid to move one grid period (1.2074~arcsec) over the reference point.

Equation~(\ref{eq:ik}) is the most general representation of the signal
produced by any object.  In order to interpret this signal in terms of
an object model it is necessary to know what the signal would be for a
point source of unit intensity; in other words, we need the equivalent
of the point-spread function in image processing.  The rectification 
mentioned in Sect.~\ref{sec:grid} meant that the expected signal from a
point source of magnitude $\mbox{\it Hp}=0$ (in the Hipparcos
photometric system) located at the reference point is:
\begin{equation}
  I_k = K[1+M_1\cos{p_k}+M_2\cos{2p_k}] \, .
  \label{eq:ik0}  
\end{equation}
Here $K=6200$~counts per sample, $M_1=0.7100$ and $M_2=0.2485$ are
constants chosen not far from the actual mean calibration values
at mid-mission (see Figs.~14.4, 14.3 and 5.6 in Vol.~3 of ESA \cite{hip}).

\subsection{Spatial frequencies}
\label{sec:freq}

As illustrated by Fig.~\ref{fig:phase} the resulting signal will depend on the
direction of scanning, even in the simple case of a point source with
a fixed position relative to the reference point.  The direction of
scanning across the reference point can be specified by a position
angle $\theta$ defined in the usual way, i.e.\ with $\theta=0$ for
a scan where the grid is moving in the direction of increasing
$\delta$ (`North'), and $\theta=90^\circ$ for a scan in the direction
of increasing $\alpha$ (`East').  The spatial frequency of the grid,
$f$ (expressed in radians of modulation per radian on the sky),
is nominally $(1296000~\mbox{arcsec})/(1.2074~\mbox{arcsec}) \simeq
1\,073\,400$.  In reality it varies slightly, mainly because of
differential stellar aberration, which changes the apparent scale
by up to $\pm 0.01$~per cent depending on the barycentric velocity 
vector of the satellite.  For the production of the TD data,
the apparent scale and orientation of the grid, as projection onto
the sky in the vicinity of the reference point, were strictly 
calculated for each transit, taking into account aberration as well
as the satellite attitude and the calibrated field-to-grid
geometric transformation.  The result is expressed as the rectangular
components of the spatial-frequency vector in the tangent plane of
the sky:
\begin{equation}
  f_x = -f\sin\theta \, , \qquad f_y = -f\cos\theta \, .
  \label{eq:fxfy}
\end{equation}
The minus signs mean that the spatial-frequency vector ($\vec{f}$)
points in the opposite direction to the motion of the grid across
the reference point.  This is purely a matter of convention, which was
adopted for historic reasons.

\subsection{The phase of an arbitrary point source}
\label{sec:source}

Consider now a point source of magnitude {\it Hp} which, in a particular
scan, is displaced by $(x,y)$ from the reference point,
where $x$ is measured positive towards increasing $\alpha$ and 
$y$ towards increasing $\delta$.  Given the spatial frequency
components of the scan, $f_x$ and $f_y$, the expected signal is:
\begin{eqnarray}
  I_k &=& 10^{-0.4{\it Hp}}K [1 + M_1\cos(p_k+\phi) + 
  M_2\cos 2(p_k+\phi)] \nonumber \\
  \label{eq:ikphi}
\end{eqnarray}
where
\begin{equation}
  \phi = f_x x + f_y y
  \label{eq:phi}
\end{equation}
is the phase shift on the grid caused by the positional offset.
[That $\phi$ is added to $p_k$ in Eq.~(\ref{eq:ikphi}), rather than
subtracted, is consistent with the sign convention adopted in the
definition of $f_x$ and $f_y$.]

Equations~(\ref{eq:ikphi})--(\ref{eq:phi}) are based on a linearization
of the transformation between spherical coordinates and their projection
on the tangent plane of the sky through the reference point.  Within
angular radius $\rho$ of the reference point, the neglected non-linear
terms are generally of order $\rho^3$, or $<0.1$~mas for $\rho<160$~arcsec.
Since the TD for a given target are normally confined to an area set by
the size of the IDT sensitive spot ($\sim 30$~arcsec), or a few times this
area (for multiple pointings), the linearization is always adequate.

The positional offset $(x,y)$ is in general caused by
a time-dependent combination of the differences between the astrometric
parameters of the star and of the reference point.  The differences in
$\alpha$, $\delta$, $\mu_{\alpha*}$ and $\mu_\delta$ are easily converted
to an offset which is a linear function of time.%
\footnote{Following the convention of the Hipparcos Catalogue, an 
asterisk ($*$) is used to denote a true arc, as in 
$\mu_{\alpha*}\equiv\mu_{\alpha}\cos\delta$.}
The parallax
difference produces a shift which is more complicated to calculate, as
it depends on the position ($\vec{s}$) of the satellite relative to the
solar system barycentre at the time of observation.  To facilitate this
calculation, a third spatial frequency component, $f_p$, is supplied
with the TD.  This is basically just the scalar product
$-\vec{s}'\vec{f}$, where the satellite position $\vec{s}$ is in
astronomical units and $\vec{f}$ is the previously defined
spatial-frequency vector of the grid.  With this definition it is
found that a parallax difference of $\Delta\pi$ relative to the
reference point causes the additional phase shift
$\Delta\phi = f_p\Delta\pi$ in Eq.~(\ref{eq:phi}).  The complete
expression for the phase of a star with astrometric parameters
$(\alpha,\delta,\pi,\mu_{\alpha*},\mu_{\delta})$ is therefore
\begin{equation}
  \phi = f_x(\Delta\alpha{*}+t\Delta\mu_{\alpha*}) 
       + f_y(\Delta\delta+t\Delta\mu_{\delta}) + f_p\Delta\pi
  \label{eq:phitot}
\end{equation}
where $t$ is the time of the transit, expressed in (Julian) years 
from J1991.25, and $\Delta\alpha{*}=(\alpha-\alpha_0)\cos\delta_0$,
$\Delta\delta=\delta-\delta_0$, $\Delta\pi=\pi-\pi_0$, 
$\Delta\mu_{\alpha*}=\mu_{\alpha*}-\mu_{\alpha*0}$, 
$\Delta\mu_{\delta}=\mu_{\delta}-\mu_{\delta0}$  are the differences 
with respect to the astrometric parameters of the reference point,
$(\alpha_0,\delta_0,\pi_0,\mu_{\alpha*0},\mu_{\delta 0})$.

The periodicity of the grid means that, in a given scan, positional
differences that are multiples of the grid period in the direction of
scanning will not produce measurable differences in the detector signal.
This `grid-step ambiguity' is normally resolved by the combination
of scans from a variety of directions in the course of the mission.

\subsection{Changing the reference point}
\label{sec:change}

The choice of reference point for the Transit Data is arbitrary, as
long as it is sufficiently near the target for linearization errors
to be negligible ($\la 160$~arcsec; see Sect.~\ref{sec:source}).
As mentioned previously, the adopted reference point usually
corresponds to the HIC values.  The user may 
ask why the final astrometric parameters in the Hipparcos and Tycho 
Catalogues were not used instead.  There were several reasons for this.
First of all, the TD were derived directly 
from intermediate files of the NDAC reduction (Sect.~\ref{sec:TD}), 
which were compiled in their final version almost a year
before the astrometric parameters of the Hipparcos Catalogue became
available.  Secondly, most of the objects are double or multiple, and
it is not evident which point to use in such cases.  Finally, TD
are available also for cases where no valid astrometric solution is
provided in the Hipparcos Catalogue.  In those cases it would have been
necessary to use something like the HIC data anyway.
Although it would have been possible, using the formulae below, to 
transform the TD to some other, possibly more natural or desirable 
reference point, such a process would in the end have been largely 
arbitrary.  In addition, since any manipulation of the 
data always entails some risk of error, however small, it was felt 
better to leave the reference points as defined in the NDAC 
intermediate data files.  The only modification applied was the
transformation from the intermediate NDAC reference frame into the 
ICRS system, which was deemed essential; this was achieved by 
rotating the positions and proper motions of the reference points, 
rather than modifying the TD coefficients. (This explains why the
position and proper motion of the reference point never coincide
{\it exactly\/} with the HIC values.)

In some circumstances it may be desirable to change the 
reference point for a set of TD.  This is particularly relevant for the 
construction of aperture synthesis images (Sect.~\ref{sec:aper}),
where the reference point defines the centre of the image.  An
object with a poorly determined pre--Hipparcos position may be
severely offset from the centre of the reconstructed map, possibly
falling outside the map altogether or creating a false image due 
to aliasing from a position outside the map.  Not only the positional
offset, but also errors in the proper motion and parallax of the
reference point may create problems for the image reconstruction.
The effect of such errors will be a blurring of the image due to
the relative motion between the reference point and the true object.
The proper motion error produces a blurring along a straight 
line in the map while the parallax error causes an additional 
elliptical blurring.  Most objects with large proper motions
($>$~100~mas~yr$^{-1}$) or parallaxes ($>$~100~mas) have reasonable,
non-zero estimates of these quantities in the Input Catalogue, so
the blurring effect is usually not very serious for qualitative
evaluation of the images.  However, for a quantitative analysis
the effect needs to be considered.

Changing the astrometric parameters of the reference point requires
that the phase of each transit is adjusted to take into account
the apparent positional offset of the new reference point from the
old one at the time of the transit.  Fortunately, this is easily 
done by means of the spatial frequency components $f_x$, $f_y$,
$f_p$ provided for each transit.  Let
$\Delta\alpha*=(\alpha_0'-\alpha_0)\cos\alpha_0$,
$\Delta\delta=\delta_0'-\delta_0$, etc., be the (small) differences 
between the new ($'$) and the old reference point
in terms of the five astrometric parameters.  The required phase
shift for a particular transit is then given by Eq.~(\ref{eq:phitot}).
The corresponding transformation of the Fourier coefficients in
Eq.~(\ref{eq:ik}) is
\begin{eqnarray}
  b_1' &=& b_1 \, , \nonumber\\
  b_2' &=& b_2\cos\phi - b_3\sin\phi  \, ,\nonumber \\
  b_3' &=& b_2\sin\phi + b_3\cos\phi  \, ,\\
  b_4' &=& b_4\cos 2\phi - b_5\sin 2\phi  \, ,\nonumber \\
  b_5' &=& b_4\sin 2\phi + b_5\cos 2\phi  \, . \nonumber
  \label{eq:shift}
\end{eqnarray}
An example of the change of reference point is shown in 
Figs.~\ref{fig:clean} and \ref{fig:shift}.

\subsection{Multiple pointings and target positions}
\label{sec:mult}

One complication in the Hipparcos satellite operation and data
reductions was caused by double and multiple stars having
separations roughly in the range 10 to 30~arcsec.  As already mentioned,
the main Hipparcos detector had a sensitive area (IFOV) of about
30~arcsec diameter.  This area could be directed
towards any pre-defined point on the sky currently within
the telescope field of view.  The IFOV pointing, calculated from 
the real-time knowledge of the satellite attitude and the celestial 
position of the target, typically
had errors of 1 to 2~arcsec rms.  Ideally, no star should be observed while
its image was close to the edge of the IFOV, where the guiding errors
might produce a distorted signal.  For double and multiple systems with
separations less than about 10~arcsec this could be achieved
by centering the IFOV somewhere in the middle of the system, so that
all components remained within the flat-topped central part of the
IFOV sensitivity profile.  For systems with separations greater than
about 30~arcsec the individual components (or subsystems with
separations below 10~arcsec) could be observed as single stars,
again avoiding signal distortion from the IFOV edges.

However, systems with intermediate separations ($\simeq 10$ to 30~arcsec)
could not be observed without some adverse effects of the IFOV
edges.  In order to allow at least some useful astrometric information
to be extracted for such systems, each component (or subsystem)
received a separate pointing.
For example, HIP~70 and HIP~71 formed such a two-pointing system
with a separation of about 15~arcsec.  When pointing at the brighter
star (HIP~71, $Hp\simeq 8.4$), the other component (HIP~70, $Hp\simeq 10.6$)
would be just at the edge of the IFOV and a (variable) fraction of its
signal was added to that of the brighter star.  Conversely, when
pointing at the fainter component, some fraction of the brighter
star's signal would be added.  Proper reduction of
such systems must consider the mutual (and possibly distorted) influence
of each component upon the other, as was indeed done in the Hipparcos data
reductions.  In some multiple systems three 
different pointings were needed.

The situation is further complicated by the fact that the targeted
position of the IFOV was sometimes updated in the course of the mission,
usually because the original (ground-based) position was found to
be wrong by several arcsec.  Knowledge of both the original and the
updated position, and the time of updating, may then be necessary
for proper interpretation of the observations.

In order to cope with two- and three-pointing systems as well as
updated positions, the concept of `target positions' was introduced
in the TD.  For a set of TD referring to a particular double or multiple
system, target positions are defined by their offset coordinates
$(\Delta\alpha*,\Delta\delta)$ from the adopted reference point,
rounded to the nearest arcsec.  
In most cases there is just one target position, coinciding with
the reference point [offset coordinates $=(0,0)$].  For multiple-pointing 
systems there is at least one target position for each
pointing, with a different HIP number attached to each pointing.
For objects whose coordinates relative to the reference point changed
in the course of the mission, a new target position was introduced 
whenever the offset coordinates changed by more than 1~arcsec.
To within the errors of the real-time attitude determination
(normally 1--2~arcsec rms) it can therefore be assumed that the
IFOV was pointed to the specified target position.

In the TD the results from different pointings pertaining to the same
system have been collected together and expressed relative to a common
reference point.  This has been done for all systems deemed to be
`difficult' in the sense explained above, but not for very wide systems
where the mutual influence of the component signals was negligible.
The TD moreover contains the bookkeeping data necessary to calculate
the actual target positions in each transit.

\begin{table}[t]
\caption[ ]{Summary of the contents of the TD index and data files on
Disk~6 of the Hipparcos Catalogue.  Data fields in the Transit Data File
are separated by the character `$|$'.  In the header record there are 13
fields designated JH1 through JH13, etc.}
\begin{flushleft}
\begin{tabular}{ll}
\hline\noalign{\smallskip}
\multicolumn{2}{@{\hspace{0pt}}l}{\it Transit Data Index File {\rm (hip\_j.idx)}}\\
\multicolumn{2}{l}{120$\,$416 records, each of 7 bytes+CR+LF = 9 bytes}\\
\multicolumn{2}{l}{The $n$th record in this file points to the header record} \\
\multicolumn{2}{l}{for HIP $n$ in hip\_j.dat (Fig.~2)} \\
\noalign{\smallskip}
\hline\noalign{\smallskip}
\multicolumn{2}{@{\hspace{0pt}}l}{\it Transit Data File {\rm (hip\_j.dat)}}\\
\multicolumn{2}{l}{4$\,$351$\,$156 records, each of 125 bytes+CR+LF = 127 bytes}\\
\multicolumn{2}{l}{The data for a given entry consist of one header record,}\\
\multicolumn{2}{l}{one pointing record, and $N_T$ transit records.}\\
\noalign{\smallskip}
\multicolumn{2}{@{\hspace{0pt}}l}{\it Header Record:}\\
JH1--3   & HIP identifiers (JH2--3~=~0 if not used) \\
JH4      & $N_P$~=~number of target positions \\
JH5      & $N_T$~=~number of transit records \\
JH6--10  & $\alpha_0$, $\delta_0$, $\pi_0$, $\mu_{\alpha*0}$, 
$\mu_{\delta 0}$~=~reference point \\
JH11--13 & assumed colour indices ($V\!-\!I$) for JH1--3 \\
\noalign{\smallskip}
\multicolumn{2}{@{\hspace{0pt}}l}{\it Pointing Record:}\\
JP1      & index (1 to 3) to JH1--3 for target position 1 \\
JP2      & offset in $\alpha$ for target position 1 (arcsec)\\
JP3      & offset in $\delta$ for target position 1 (arcsec)\\
JP4--6   & same as JP1--3 but for target position 2 \\
$\cdots$ & $\cdots$ \\
JP25--27 & same as JP1--3 but for target position 9 \\
\noalign{\smallskip}
\multicolumn{2}{@{\hspace{0pt}}l}{\it Transit Record:}\\
JT1      & $I_P$~=~target position (1 to $N_P$) for the transit \\
JT2      & $t$~=~epoch of the transit (years from J1991.25)\\
JT3--5   & $f_x$, $f_y$, $f_p$~=~spatial frequencies \\
JT6      & $\ln b_1$ \\
JT7--10  & $b_i/b_1$, $i=2\dots 5$ \\
JT11--15 & $\ln\sigma_i$, $i=1\dots 5$ (standard errors of $b_i$) \\
JT16--17 & $s_1$, $s_2$~=~colour correction factors \\
JT18     & $\sigma_{\rm att}$~=~additional attitude noise (mas)\\
JT19     & computed (0) or assumed (1) standard errors \\
\noalign{\smallskip}
\hline
\end{tabular}\label{tab:sum}
\end{flushleft}
\end{table}

\section{The Transit Data files}
\label{sec:files}

This section gives a rather detailed description of the contents of
the TD files, complementing the formal description in Vol.~1 of ESA
(\cite{hip}) and providing additional explanation of the data items.  All
relevant data are contained in two ASCII files, both located on
Disk~6 in Vol.~17 of ESA (\cite{hip}): the TD index file (hip\_j.idx,
$\simeq 1$~Mb) and the TD file (hip\_j.dat, $\simeq 553$~Mb).
A summary of the contents of the two files is in Table~\ref{tab:sum}, 
while important relations among the data are illustrated in
Fig.~\ref{fig:idx}.

\subsection{The index file: hip\_j.idx}
\label{sec:index}

The TD index file is included to facilitate accessing the TD file.  It
contains a pointer from the HIP number to the corresponding record in
the TD file where data on that object can be found.  Since HIP numbers
range from 1 to 120416, there are exactly 120416 records in the index
file.  Each record consists of a 7-character integer, which is the
record number in the TD file for the relevant header record
(Sect.~\ref{sec:header}).  For example, TD for the double star
HIP~7 can be accessed by first reading record number 7 in the
index file (hip\_j.idx).  The content of that record is the
integer 129.  Record number 129 in the TD file (hip\_j.dat)
is thus the header record for HIP~7, and information on that object
is contained in that and subsequent records of the TD file.  If no TD
are available for a given HIP number, then the corresponding index file
record contains the number $-1$.

\begin{figure}[t]
  \resizebox{\hsize}{!}{\includegraphics*{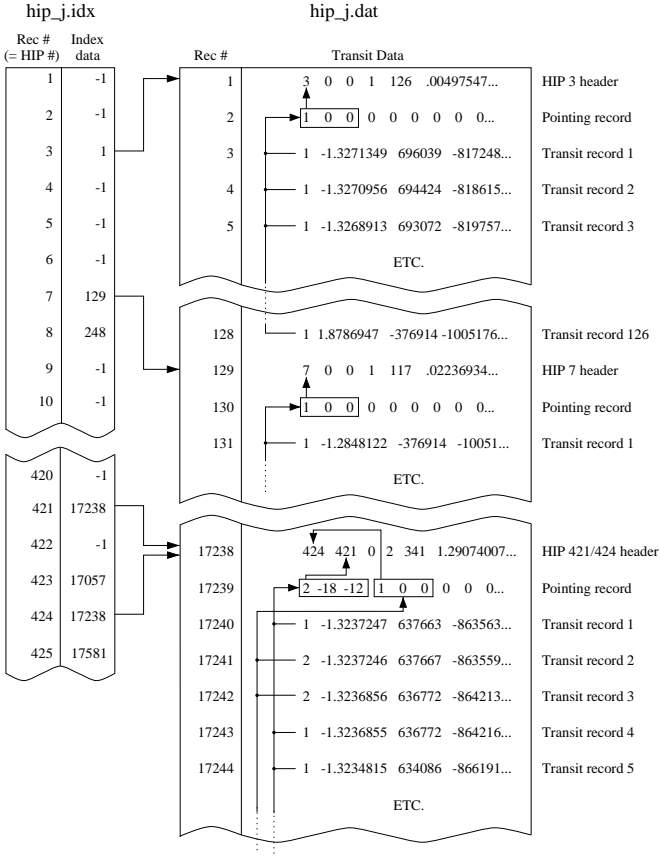}}
  \caption[ ]{Relations among the contents of the index and data files.}
\label{fig:idx}
\end{figure}

For two- and three-pointing systems (Sect.~\ref{sec:mult}) there are
two or three different index file entries pointing to the same
TD record.  An example is shown in Fig.~\ref{fig:idx}, where the
index file entries for HIP~421 and HIP~424 both point
to the 17238th record in the TD file.

For efficient accessing of the TD it is recommended that the index file
is read into computer prime memory as a one-dimensional integer array of
length 120416.  This array is then used as a look-up table for finding
data in the TD file.  The TD file has a constant record length of
127~bytes (including the two end-of-record characters
$\backslash$r$\backslash$n~=~CR+LF), which allows simple direct access
to any given record number.

\subsection{The TD file: hip\_j.dat}
\label{sec:dat}

\subsubsection{The header record}
\label{sec:header}

The header record is the first record in the TD file containing data on
a specific system.  The index file always points to this header record
for a given HIP number.  The header record contains general information
about the system in 13 data fields.  In holding to the Hipparcos Catalogue conventions,
the 13 data fields in the header record will be
called JH1, JH2, $\dots\;$, JH13.

The first three data fields (JH1--3) contain the HIP numbers relevant to
the subsequent transit records.  JH1 is always defined; JH2 and JH3
are only defined for two- and three-pointing systems (Sect.~\ref{sec:mult}).
The number of pointings, or more accurately the number of different
target positions ($N_P$), is specified in field JH4.  Details
on the relative pointings are given in the pointing record
(Sect.~\ref{sec:point}).

JH5 gives the number of transits ($N_T$), or observations, for a given
system.  Since there is one record for each transit, this is also
the number of transit records.  The total number of
records for an object is then $N_T+2$, where the header
and pointing records make up the extra two records.

JH6--JH10 give, in order, the right ascension (deg), declination (deg),
parallax (mas), proper motion in right ascension (mas~yr$^{-1}$) and proper
motion in declination (mas~yr$^{-1}$) of the reference point.  The Fourier
coefficients $b_1$--$b_5$ of the subsequent transit records are
expressed relative to this reference point. 
It is important to realize that these values are usually derived from
the Input Catalogue and therefore do not agree with the astrometric
parameters given in the Hipparcos Catalogue.  For instance, the
parallax of the reference point is often zero.
The position refers to the epoch J1991.25, which was close to mid
mission.  The reference system is the same as that of the Hipparcos
Catalogue, viz.\ the International Celestial Reference
System (ICRS) (Feissel \& Mignard \cite{icrs}).

JH11--JH13 hold the assumed colour index $V\!-\!I$ (mag) for each of the
three HIP numbers in JH1 through JH3.  This value $V\!-\!I$ was used
in rectifying the signals from each transit.

\subsubsection{The pointing record}
\label{sec:point}

The pointing record defines the offset of the target position
of each subsequent transit with respect to the reference point.
The pointing record may define several different target positions
as required for multiple-pointing objects. The actual number of 
different target positions ($N_P$) is given in JH4 of
the header record.  Each target position is specified by three numbers: the
first takes the value 1, 2 or 3, depending on which of the HIP numbers in JH1,
JH2 or JH3 that the target position refers to.  The second and third numbers
give the offset in $\alpha$ and $\delta$, respectively, of the target
position from the reference point.  These last two values are rounded
to the nearest arcsecond.

Because every record of the TD file has 125 bytes, this allowed up
to nine different target positions to be defined in the pointing record.
In reality the maximum $N_P$ was 7.

Most objects used a single pointing, and the astrometric values in the
Input Catalogue were sufficiently accurate that no updating was required.
Moreover, the Input Catalogue values were usually taken as the reference
point for the transit data.  In this case $N_P=1$ and the first and only
entry in the pointing record contains the three values:  `1 0 0' (cf.\ the
excerpts for HIP~3 and 7 in Fig.~\ref{fig:idx}).  The `1' refers to the HIP
number in JH1.  The `0 0' means that the target position coincided 
(to the nearest arcsec) with the reference point.  Subsequent transit
records have `1' in the first field (JT1) indicating that the target
position for each transit was as given in the first entry of the
pointing record.

In the trivial case of a single, well-centered pointing, as illustrated
by HIP~3 and 7 above, the pointing record is really not needed.  However, for 
those objects with multiple HIP entries and pointings, the pointing data
provide important information.  This is exemplified by the excerpt
for the two-pointing system HIP~421$+$424 shown in the lower part of 
Fig.~\ref{fig:idx}.  As indicated in the header record, there are 341
transits for this system, made using two different target positions.
The first transit for this system (in record 17240)
was made with the IFOV pointing at the first target position 
(`2~~--18~~--12'),
i.e.\ referring to the second HIP entry (421) and pointing 18~arcsec
to the west and 12~arcsec to the south of the reference point.  The
next transit (record 17241) was made in the second pointing (`1 0 0'),
i.e.\ referring to the first HIP entry (424) and centred on the reference
point.  The third transit (record 17242) was also made in the second
pointing, while the fourth transit was made in the first pointing, and
so on.

The fact that Hipparcos had two fields of view, separated by the `basic
angle' of $58^\circ$, is largely irrelevant for the use of the TD.  There
is in fact no flag in the TD telling in which field of view a 
particular transit occurred.

\subsubsection{The transit data record}
\label{sec:tdr}

The actual scan information is contained in the transit data record.
This section will cover the entries contained in these records, as well as
some physical interpretations.

The first field (JT1) for each record tells which of the $N_P$ (=~JH4)
target positions was used to observe the transit.  The second field (JT2)
gives the time of the observation, in years, from the epoch J1991.25.
The time is measured in Julian years of exactly 365.25 days.  The epoch
J1991.25 is equivalent to the Julian date JD 2$\,$448$\,$349.0625 on
the Terrestrial Time (TT) scale.

The next three entries (JT3 to JT5) contain the spatial frequencies $f_x$,
$f_y$ and $f_p$ defined as in Sects~\ref{sec:freq} and \ref{sec:source}.  
They are expressed in units of rad~rad$^{-1}$ (radian of modulation phase 
on the grid per radian on the sky).

The entries JT6 through JT10 contain, in coded form, the five Fourier
coefficients $b_1$--$b_5$ defined by Eq.~(1).  JT6 contains the natural
logarithm of $b_1$ ($\ln{b_1}$) and JT7 to JT10 contain the normalized 
coefficients ($b_2$/$b_1$, $b_3$/$b_1$, $b_4$/$b_1$ and $b_5$/$b_1$).
This coding was adopted because $b_1$, being the mean intensity of the 
signal, is always positive and spanning a wide range of values according to
the magnitude of the object; $b_2$ through $b_5$, on the other hand, may 
have either sign but are always numerically smaller than $b_1$.

The next five entries (JT11--15) contain the natural logarithms of
the standard errors ($\sigma_1$--$\sigma_5$) for each of the Fourier
coefficients.  JT19, the last entry in this record, is a flag for the 
standard errors.  JT19 is normally set to zero.  It is set to `1' 
if the standard errors could not be computed
in the normal way, but were estimated roughly from the standard errors
of the other transits.  A non-zero
value in JT19 is thus a warning that the standard errors in JT11--15 
should be treated with caution.

JT16--17 contain two colour correction factors, $s_1$ and $s_2$.
As previously mentioned, the TD are rectified signals, i.e.\ corrected for
calibrated variations in the geometric and photometric responses.  Among
the many factors that have been taken into account in this process are
the colour effects with respect to position in the field of view and time 
variations of the photometric calibrations.  These corrections depend on
an assumed colour for the object, as given in JH11--13.  If it should turn
out that the assumed colour for a particular transit was considerably in
error, $s_1$ and $s_2$ provide a possibility to correct (approximately)
the Fourier coefficients accordingly.  See Sect.~2.9.3 in Vol.~1 of the 
Hipparcos and Tycho Catalogues for details of the correction 
procedure.

The second to the last entry (JT18) deals with the errors in the
attitude determination.  The phase determination of each scan was
basically limited by photon noise, but the standard errors in JT11--15
also contain a contribution from the along-scan attitude uncertainty.
It turned out to be very difficult to treat the attitude
errors rigorously in the TD, and as a result these errors were generally 
underestimated.  JT18 contains an estimate of the additional attitude
noise ($\sigma_{\rm att}$, in mas) required in order
to derive, from the TD, astrometric standard errors that are consistent 
with the main processing of Hipparcos data.

\subsubsection{Known errors in the Transit Data}
\label{sec:tde}

After publication of the Transit Data CD-ROM, six cases of Fortran format 
overflow errors have been discovered in the file hip\_j.dat.  No useful 
data are lost because of the errors, but special measures may be needed
to read the corresponding records.  The interface programs described in
Sect.~\ref{sec:soft} automatically correct these errors.  Further details 
can be found at the Internet address given in Sect.~\ref{sec:soft}.

\section{Aperture synthesis imaging}
\label{sec:aper}

The Hipparcos satellite was not designed for imaging and did not contain
any imaging device such as a CCD camera.  The combination of a
modulating grid and the IDT, 
while well adapted to the observation of isolated 
point sources, was far from ideal for the observation of more 
complex resolved objects.  As explained in Sect.~\ref{sec:grid} the
grid essentially extracted two spatial frequencies (with periods
1.2074~arcsec and 0.6037~arcsec in the direction of each scan)
of whatever intensity distribution was within the 30~arcsec sensitive
spot.  This is much more reminiscent of spatial interferometry
than of normal optical imaging.  In fact, the two harmonics of the
detector signal correspond to the fringes produced by an object in
two interferometers with baselines of 94~mm and 188~mm, respectively
(assuming an effective wavelength of 550~nm).  Image reconstruction techniques
using interferometric observations have for a long time been standard
in the radio astronomical community.  It was therefore quite natural to 
apply these techniques to the Hipparcos data 
(Lindegren \cite{lind}; Quist et al.\ \cite{quist}).

\subsection{Mathematical foundation} 
\label{sec:math}

Equation~(\ref{eq:ikphi}) gives the signal for a point source 
located at the position $\vec{r}=(x,y)$ relative the reference point.
Now consider an extended object with the general brightness 
distribution $B(\vec{r})$.  Summing up the contributions to the
detector signal from each element of the sky we find (apart from a
constant scaling factor)
\begin{eqnarray}
  I(p) &=& \int\!\!\!\int S(\vec{r})B(\vec{r}) 
           [ 1 + M_1\cos(p+\vec{f}\cdot\vec{r}) \nonumber \\ 
        && \qquad {}+ M_2\cos(2p+2\vec{f}\cdot\vec{r})]\,{\rm d}^2\vec{r}
  \label{eq:ip}
\end{eqnarray}
where $S(\vec{r})$ is the sensitivity profile of the instantaneous 
field of view.  Introducing the Fourier transform
\begin{equation}
  V(\vec{f}) = \int\!\!\!\int S(\vec{r})B(\vec{r})
               \exp({\rm i}\vec{f}\cdot\vec{r})\,{\rm d}^2\vec{r}
  \label{eq:vf}
\end{equation}
we find that Eq.~(\ref{eq:ip}) can be written
\begin{eqnarray}
  I(p) &=& V(\vec{0}) + \mbox{Re}[V(\vec{f})]M_1\cos p
         - \mbox{Im}[V(\vec{f})]M_1\sin p \nonumber\\
       &+& \mbox{Re}[V(2\vec{f})]M_2\cos 2p
         - \mbox{Im}[V(2\vec{f})]M_2\sin 2p \, .
  \label{eq:ip0}
\end{eqnarray}
Comparison with Eq.~(\ref{eq:ik}) shows that
\begin{eqnarray}
  V(\vec{0}) &=& b_1 \, , \nonumber \\
  V(\vec{f}) &=& (b_2-\mbox{i}b_3)/M_1 \, , \quad
  V(-\vec{f}) = V^\ast(\vec{f}) \, , \nonumber \\
  V(2\vec{f}) &=& (b_4-\mbox{i}b_5)/M_2 \, , \quad
  V(-2\vec{f}) = V^\ast(2\vec{f}) \, ,
  \label{eq:v0}
\end{eqnarray}
where the asterisk denotes the complex conjugate.
Since $M_1$ and $M_2$ are conventional constants (Sect.~\ref{sec:phase})
it is seen that a single transit defines the complex function $V$ in the
five points $\vec{0}$, $\pm\vec{f}$, $\pm 2\vec{f}$, of the spatial
frequency plane.  However, the conjugate symmetry of $V$ means that 
there are only three independent complex visibilities per transit.
Successive transits of the same object are made at different spatial
frequencies $\vec{f}=(f_x,f_y)$ and in the course of the mission
knowledge of the function $V$ is built up in a number of 
different points.  To the extent that $S(\vec{r})B(\vec{r})$
remains constant over the mission, it may then be recoverable from
$V$ using standard image reconstruction techniques.

In the context of radio interferometry and aperture synthesis, we
may identify $V(\vec{f})$ with the complex visibility function
associated with the source brightness distribution $B(\vec{r})$
and single-antenna reception pattern $S(\vec{r})$ 
(Thompson et al.\ \cite{thom}).  The visibility 
function is usually expressed in terms of coordinates $(u,v)$ 
which give the projection of the interferometer baseline on the sky 
plane and are expressed in wavelengths.  The relation to the TD 
spatial frequency components is simply
\begin{equation}
  u=f_x/2\pi \, , \quad v=f_y/2\pi \, .
  \label{eq:uv}
\end{equation}
Our reference point is equivalent to the phase reference position
used in connected-element radio interferometry 
(Thompson et al.\ \cite{thom}), or to the strong reference point source 
(typically a quasar) used in phase-referenced VLBI observations 
(Lestrade et al.\ \cite{lestr}).

\begin{figure}[t]
  \resizebox{\hsize}{!}{\includegraphics*{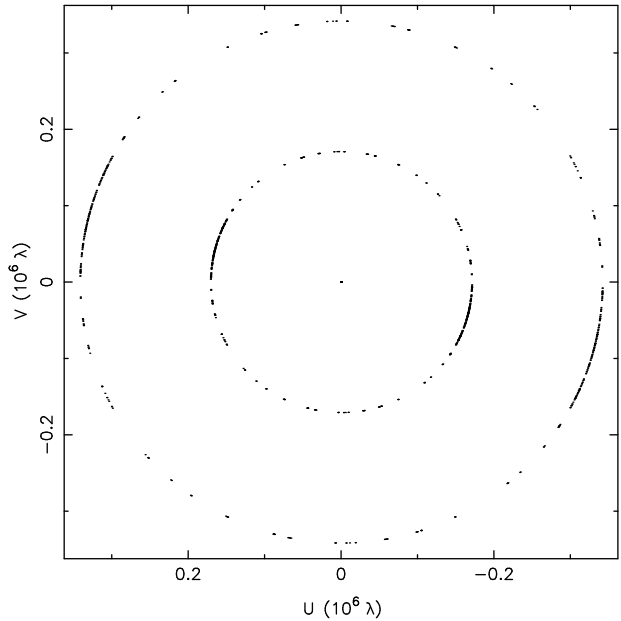}}
  \caption[ ]{UV coverage for HIP~97237.  For every point $(u,v)$ marked in
this diagram there is a measurement of the complex visibility $V$ representing
the amplitude and phase of the detector signal during a certain transit.  Each 
transit provides five equidistant visibilities (including the
origin) oriented along a straight line through the origin.  The celestial
orientation of the line is normal to the slits of the Hipparcos modulation grid
at the transit.  $u$ is the spatial frequency in the east--west direction,
expressed in modulation periods (`wavelengths') per radian on the sky;
$v$ is the north--south component of the spatial frequency in the same unit.}
\label{fig:uv}
\end{figure}

The distribution of the observations in the $uv$ plane is
all-important for the possibility to reconstruct complicated images
from the measured visibilities $V(u,v)$.  Unfortunately the Hipparcos 
scanning law and the use of a modulating grid with just a single period
seriously limit the $uv$ coverage of the TD.  According to 
Eqs.~(\ref{eq:fxfy}) and (\ref{eq:uv}) the coverage is limited to
the central point $(u,v)=(0,0)$ and two concentric rings with
radii $\simeq 170\,830$ and $341\,660$ wavelengths.  Moreover, for
objects in the ecliptic region of the sky (ecliptic latitude
$|\beta| \la 45^\circ$) the scanning law constrains the scan
angle $\theta$ such as to produce a gap of `missing' scans roughly
in the east--west direction.  At $|\beta| \simeq 47^\circ$ there is
instead a surplus of scans in the east--west direction (cf.\ 
Fig.~\ref{fig:uv}).  For high-latitude objects, finally, the coverage 
is usually more uniform in $\theta$.
 
In continuing the analogy with radio interferometry, we will discuss
in the next section how images can be produced using the Transit Data.
Utilizing the experience developed for aperture synthesis imaging, we
use only publicly available software for producing and deconvolving
these images.

\subsection{Conversion of TD to UV-FITS format}
\label{sec:tr2uv}

The images presented here were produced using the Caltech Difmap
software package (Shepherd \cite{shep}).  Another software package that 
could be used for the analysis is AIPS (Astronomical Image
Processing System) developed at the National Radio Astronomy 
Observatories (NRAO).  Both Difmap and
AIPS take input data in the form of FITS files, using the `Random
Groups' format (NOST \cite{nost2}).  This amendment to the basic FITS 
format is more commonly called the UV-FITS format, since it is used 
almost solely for interferometry data.

We have written and made publicly available (Sect.~\ref{sec:soft})
a concise Fortran program which reads TD from the CD-ROM format 
(hip\_j.idx and hip\_j.dat) for any given object and produces an 
output file in the UV-FITS format.
In the following we describe some of the features
of the UV-FITS format and how the TD were adapted to it.

The basic FITS (Flexible Image Transport System; 
Wells et al.\ \cite{fits1}; NOST \cite{nost1}),
well known in optical astronomy, was designed to transport digital
data in the form of $n$-dimensional regular arrays, such as CCD images,
with associated information on coordinates, dates, scales, units, 
etc.\ given in an ASCII header.  However, aperture synthesis visibility data
do not come in regular arrays, at least not in all axes, and thus
an amendment to the basic FITS was required allowing the definition of 
ordered sets of small arrays (Greisen \& Harten \cite{fits2}).
In the following we assume that the reader
has some rudimentary knowledge about the basic FITS format.

In a UV-FITS file each random group contains the visibility data
associated with a particular point in $uv$ space and time.  The
group consists of a set of parameters followed by a regular array
of measurements.  The parameters are, for instance, the $uv$ 
coordinates and date of the measurements.  The measurement array
may be multi-dimensional with, for instance, the different frequencies
and Stokes (polarization) components marked along two of the axes.
In the UV-FITS header the use of random groups is signified by
having a first axis of length zero (${\tt NAXIS1}=0$) and by
setting the keyword ${\tt GROUPS}={\tt T}$ (true).  Visibility
data are stored as three values in the measurement array, namely 
the real part of the visibility, the imaginary part, and an 
associated weight.  In the array, this corresponds to an axis of 
length 3 and type ${\tt COMPLEX}$.

For the Hipparcos TD there are three visibility measurements
per transit, corresponding to the spatial frequencies
$\vec{0}$, $\vec{f}$ and $2\vec{f}$.  The total number of
random groups is therefore ${\tt GCOUNT}=3N_T$.
${\tt PCOUNT}=6$ parameters specify each group, namely
the Fourier coordinates $(u,v,w)$, the baseline (defining which
pair of antennae that formed the interferometer), and the 
date of the observation (split in two numbers containing the 
integer and fractional parts of the Julian date).  For the
TD the $w$ coordinate is always zero.

The UV-FITS format requires that the $(u,v)$ coordinates are
expressed in seconds, while in Eq.~(\ref{eq:uv}) they are 
dimensionless.  The conversion factor requires the specification
of a reference wavelength, for which we arbitrarily adopted 
$\lambda_0=550$~nm.  The corresponding
scale factor for the $(u,v)$ coordinates in Eq.~(\ref{eq:uv})
is then ${\tt PSCAL1}={\tt PSCAL2}=\lambda_0/c$, where $c$ is
the speed of light.  For consistency, the frequency associated
with each TD observation must then be given as $c/\lambda_0$.

The measurement array in each group is 5-dimensional 
(${\tt NAXIS}=6$, since the first axis has zero length for
group data).  Its size is
${\tt NAXIS2}\times{\tt NAXIS3}\times{\tt NAXIS4}\times
{\tt NAXIS5}\times{\tt NAXIS6}=3\times 1\times 1\times 1\times 1=3$,
where the axes are of type ${\tt COMPLEX}$, ${\tt STOKES}$,
${\tt FREQ}$, ${\tt RA}$ and ${\tt DEC}$, respectively.  The values
on each axis are specified in the header by ${\tt CRVAL}n$ for
$n=2\dots 6$; in particular the frequency is given by
${\tt CRVAL4}=c/\lambda_0$ and the reference position by
${\tt CRVAL5}=\alpha_0$ and ${\tt CRVAL6}=\delta_0$.
The three values in the measurement array are, as mentioned 
before, the real and imaginary parts of $V(u,v)$ and an associated 
weight.  For the TD the weight is always equal to $1$.

The aperture synthesis programs also require the names and
geocentric positions of the antenna stations to be specified,
although this is rather pointless in our case.  We formally 
specify six stations and identify a different pair with each 
spatial frequency.  The stations are arbitrarily named to form 
the acronym `HIPUVF', which will appear on some plots.

An apparent limitation of the FITS format is the lack of keywords
for proper motion and parallax.  Until such keywords become standard, 
the proper motions and parallax values of the reference point
are given in the ASCII header as comments.

\subsection{Example images}
\label{sec:images}

Once the data are converted into a suitable form, images are quickly
produced using standard aperture synthesis programs.  Using the Difmap 
package from Caltech (Sect.~\ref{sec:soft}), we give here as an example 
a reduction of the TD for HIP~97237.  In the Hipparcos Catalogue, no
astrometric solution is given for this relatively faint ($V\simeq 12.4$) 
object.  In the Hipparcos Input Catalogue it is
noted as a double star (CCDM 19458+2707) with separation 0.9~arcsec and 
component magnitudes 12.7 and 13.6.  The ecliptic latitude of the 
object is $\beta \simeq +47^\circ$.  The reference point for this object 
is 
$\alpha_0=296.43849356$~deg,
$\delta_0=+27.12735771$~deg,
$\pi_0=97.00$~mas,
$\mu_{\alpha * 0}=-25.43$~mas~yr$^{-1}$,
$\mu_{\delta 0}=-1228.86$~mas~yr$^{-1}$.

Figure~\ref{fig:uv} shows the UV-coverage of HIP~97237.  It is seen that 
the object received rather many scans, although predominantly in the 
(ecliptic) east--west direction.  The `dirty beam' (Fig.~\ref{fig:dbeam})
reflects this anisotropy as a characteristic pattern along the east--west
section, reminiscent of the basic light modulation curve in 
Fig.~\ref{fig:phase}.

\begin{figure}[t]
  \caption[ ]{The dirty beam (point spread function) associated with the 
UV coverage of HIP~97237 shown in Fig.~\ref{fig:uv}.}
\label{fig:dbeam}
\end{figure}

\begin{figure}[t]
  \caption[ ]{The dirty map of HIP~97237, obtained by an inverse Fourier
transform of all the measured complex visibilities.}
\label{fig:dmap}
\end{figure}

The `dirty map', basically obtained as the inverse Fourier transform of 
the complex visibilities, is shown in Fig.~\ref{fig:dmap}.  Already from 
this image it is obvious that the object was offset from the expected 
(HIC) position by about 5~arcsec.  This is probably the reason 
why an acceptable solution was not found in the Hipparcos astrometric 
reductions.  Deconvolution of the dirty image, using the dirty beam as
kernel (point spread function), can be achieved by means of the CLEAN
algorithm (H{\"o}gbom \cite{clean}) implemented in Difmap.  One result of this
process (which depends on several parameters selectable in Difmap) is
shown in Fig.~\ref{fig:clean}.  Both components of the double star are
now clearly seen.  The offset of the primary component from the 
reference point, estimated from the cleaned image, is 
$(\Delta\alpha*,\Delta\delta) \simeq (+3.6,+3.6)$~arcsec.  The position 
of the secondary component relative to the primary is approximately 
0.9~arcsec towards position angle $330^\circ$.  The power in the primary 
peak of the cleaned map is 0.0455~units, where 6200 units corresponds to 
magnitude $Hp=0$ (Sect.~\ref{sec:phase}); the magnitude of the primary can
thus be estimated at $Hp \simeq 12.8$.

Figure~\ref{fig:shift} illustrates the change of reference point for the
TD described in Sect.~\ref{sec:change}.  The astrometric parameters of
the primary component in HIP~97237, relative to the reference point, were 
estimated by means of the model fitting procedure described in 
Sect.~\ref{sec:fit}.  The approximate results were 
(cf.\ Fig.~\ref{fig:output}) $\Delta\alpha*=+3603$~mas, 
$\Delta\delta=+3640$~mas, $\Delta\pi=-16$~mas, 
$\Delta\mu_{\alpha*}=-50$~mas~yr$^{-1}$,   
$\Delta\mu_\delta=-25$~mas~yr$^{-1}$.  Applying the corresponding
phase shifts to the TD, according to Eqs.~(\ref{eq:phitot}) and 
(\ref{eq:shift}),
effectively changes the reference point to coincide with the primary
component.  Performing the image synthesis on the modified TD gives
the cleaned image in Fig.~\ref{fig:shift}.  As expected, the primary 
now appears at the centre of the map.  The shift in parallax and
proper motion of the reference point improves the relative phasing
of the superposed scans, resulting in a slightly increased peak power 
(from 0.0455 to 0.0457~units).

\begin{figure}[t]
  \caption[ ]{The cleaned map of HIP~97237, obtained by deconvolution of
Fig.~\ref{fig:dmap} with the beam in Fig.~\ref{fig:dbeam}.  The cleaned
map clearly reveals the two components of the double star.}
\label{fig:clean}
\end{figure}

\begin{figure}[t]
  \caption[ ]{The cleaned map of HIP~97237 resulting from modified TD,
in which the astrometric parameters of the reference point were shifted
to coincide with the astrometric parameters of the primary component.
determined using the model fitting procedure described 
in Sect.~\ref{sec:fit}.}
\label{fig:shift}
\end{figure}

\section{Model fitting}
\label{sec:fit}

The aperture synthesis imaging attempts to reconstruct the brightness 
distribution on the sky, in principle without making any a~priori
assumption about the object.  This is excellent for exploring cases
where the nature of the object is uncertain, e.g.\ concerning the number 
of resolved components in a multiple star or their approximate positions.
The method is less useful for accurate quantitative evaluation, in
particular because the offsets in parallax and proper motion merely produce
a blurring of the image.  Since the objects of interest here consist of
a small number of point sources, direct modeling of the TD in terms of 
simple superposed signal components is usually possible.  Such model 
fitting provides the most direct and accurate estimates of specific object
parameters such as the trigonometric parallax or orbital elements.
In this section we outline the fitting procedure and give an
example of its practical realization by means of a publicly 
available computer program.

\subsection{General method}
\label{sec:genfit}

Let us assume that the object consists of $n$ point sources with intensities
$A_j$ and positions $x_j$, $y_j$ relative the reference point ($j=1 \dots n$).  
In general $A_j$, $x_j$, $y_j$ vary with time, and so may be different for the
different transits of the same object.  Given a specific model of the object
we express $A_j$, $x_j$, $y_j$ as functions of time $t$ and a set of 
model parameters $\vec{a}$.  For instance, in the 
case of a non-variable orbital binary, $\vec{a}$ would consist of 15 parameters, 
viz.\ the five astrometric parameters of the mass centre, the magnitude of each 
component, the mass ratio, and seven elements for the relative orbit.  Generally 
speaking, the object model is thus completely specified by $n$ and the functions 
$A_j(t,\vec{a})$, $x_j(t,\vec{a})$, $y_j(t,\vec{a})$ for $j=1 \dots n$.  In the
equations below we suppress, for brevity, the explicit dependence on $t$ and 
$\vec{a}$.

For a given transit the expected signal is modeled as the sum of the signals 
from the individual components, using Eqs.~(\ref{eq:ikphi}) and (\ref{eq:phi}).
Thus,
\begin{eqnarray}
  I_k &=& \sum_{j=1}^n A_j [ 1 + M_1\cos(p_k+f_xx_j+f_yy_j) \nonumber \\
  && \qquad {} + M_2\cos 2(p_k+f_xx_j+f_yy_j) ] \, .
\end{eqnarray}
Expanding the trigonometric functions and equating the terms with those in
Eq.~(\ref{eq:ik}) yields
\begin{eqnarray}
  b_1 &=& \phantom{-M_1}\sum_j A_j \, , \nonumber \\
  b_2 &=& \phantom{-}M_1\sum_j A_j \cos(f_xx_j+f_yy_j) \, , \nonumber \\
  b_3 &=&           -M_1\sum_j A_j \sin(f_xx_j+f_yy_j) \, ,\label{eq:phaseelem} \\
  b_4 &=& \phantom{-}M_2\sum_j A_j \cos 2(f_xx_j+f_yy_j) \, , \nonumber \\
  b_5 &=&           -M_2\sum_j A_j \sin 2(f_xx_j+f_yy_j) \, . \nonumber
\end{eqnarray}
Recall that $A_j$, $x_j$, $y_j$ depend on the model parameters $\vec{a}$.
The general procedure is then to adjust $\vec{a}$ in such a way that, for
the whole set of transits, the calculated signal parameters $b_1$--$b_5$ 
from Eq.~(\ref{eq:phaseelem}) agree, as well as possible, with the observed 
values.  The adjustment may use the weighted least-squares method, using 
the standard errors of the observed signal parameters to set the weights; 
but other (and more robust) metrics can also be used.  In general the
problem can be formulated as a constrained minimization problem in the 
multi-dimensional model parameter space.

The trigonometric functions in Eq.~(\ref{eq:phaseelem}) mean that the signal
parameters $b_i$ depend in a highly non-linear manner on the model parameters
which affect $x_j$ and $y_j$.  For instance, in terms of a displacement of 
one of the point sources, the effect on $b_4$ and $b_5$ is approximately 
linear only for displacements less than about $1/2f \simeq 0.1$~arcsec,
corresponding to 1~rad change in the modulation phase.  In the aperture
synthesis imaging this non-linearity is manifest in the complex structure
of the `dirty beam' (Fig.~\ref{fig:dbeam}) at all spatial scales larger 
than about 0.1~arcsec.  Additional non-linearities in the complete object 
model may result from the geometrical description of the source positions,
e.g.\ in terms of orbital elements.

The non-linearity of the object model has two important consequences for the
model fitting.  Firstly, it is usually necessary to use a non-linear, iterative 
adjustment algorithm, such as the Levenberg--Marquardt method 
(Press et al.~\cite{nr2}).  Secondly, a good initial guess of the model 
parameters
is usually required.  In particular the parameters directly affecting the
positions of the point sources need to be specified to within (what 
corresponds to) a few tenths of an arcsec.  Without a good initial guess,
the adjustment algorithm is likely to converge on some local minimum,
typically resulting in positional errors of (approximately) an integer number
of grid periods.  The correct solution, corresponding to the global minimum,
may in principle always be found through sufficiently extensive searching of 
the parameter space.  Alternatively, sufficiently good initial guesses of the 
point source positions can often be obtained from the aperture synthesis 
imaging.

Various least-squares model fitting procedures were used for the reduction of
double and multiple stars during the construction of the Hipparcos Catalogue
(see Mignard et al.\ \cite{mign} and references therein).  The double-star 
processing of the NDAC data reduction consortium 
(S{\"o}derhjelm et al.\ \cite{ss92}) essentially used the technique outlined 
above, taking the so-called Case History Files (a precursor to the TD) as input.

Perhaps the greatest potential of the TD lies in the possibility to combine
the Hipparcos data with independent observations from other instruments and
epochs.  For instance, full determination of a binary orbit generally
requires data covering at least a whole period.  Ground-based speckle
observations can sometimes provide this, constraining the geometry of the 
relative orbit much better than the Hipparcos data alone, and in turn 
leading to a better-determined space parallax.  In some favourable cases
the location of the mass centre in the relative orbit (and hence the mass
ratio) can be determined (S{\"o}derhjelm et al.\ \cite{ss97};
S{\"o}derhjelm \cite{ss99}).

One complication of the Hipparcos double star processing has been the wide
variety of applicable object models, and the consequent need to experiment
and interact with the solutions.  This process may be much facilitated by using
general and flexible software for the model fitting, rather than highly
specialized routines.  An example of this is given below.

\subsection{Model fitting using GaussFit}

GaussFit (Jefferys et al.\ \cite{gf}, \cite{gfman}) 
is a general program for the solution of
least squares and robust estimation problems, developed as a platform
to facilitate astrometric reduction of data from the Hubble Space Telescope.
It is written in the C programming language and may thus be run under a variety
of operating systems.  In this section we outline the use of GaussFit for 
model fitting to the TD, again using the binary HIP~97237 as illustration.

GaussFit was used by S{\"o}derhjelm (\cite{ss99}) in a systematic 
re-examination of 
the solutions for several double and multiple objects, through a combination
of TD with ground-based observations.  Although not illustrated in the example
below, the introduction of additional data (e.g.\ relative positions from
speckle observations) is quite straightforward by means of GaussFit.

\begin{figure}[t]
  \resizebox{\hsize}{!}{\includegraphics*{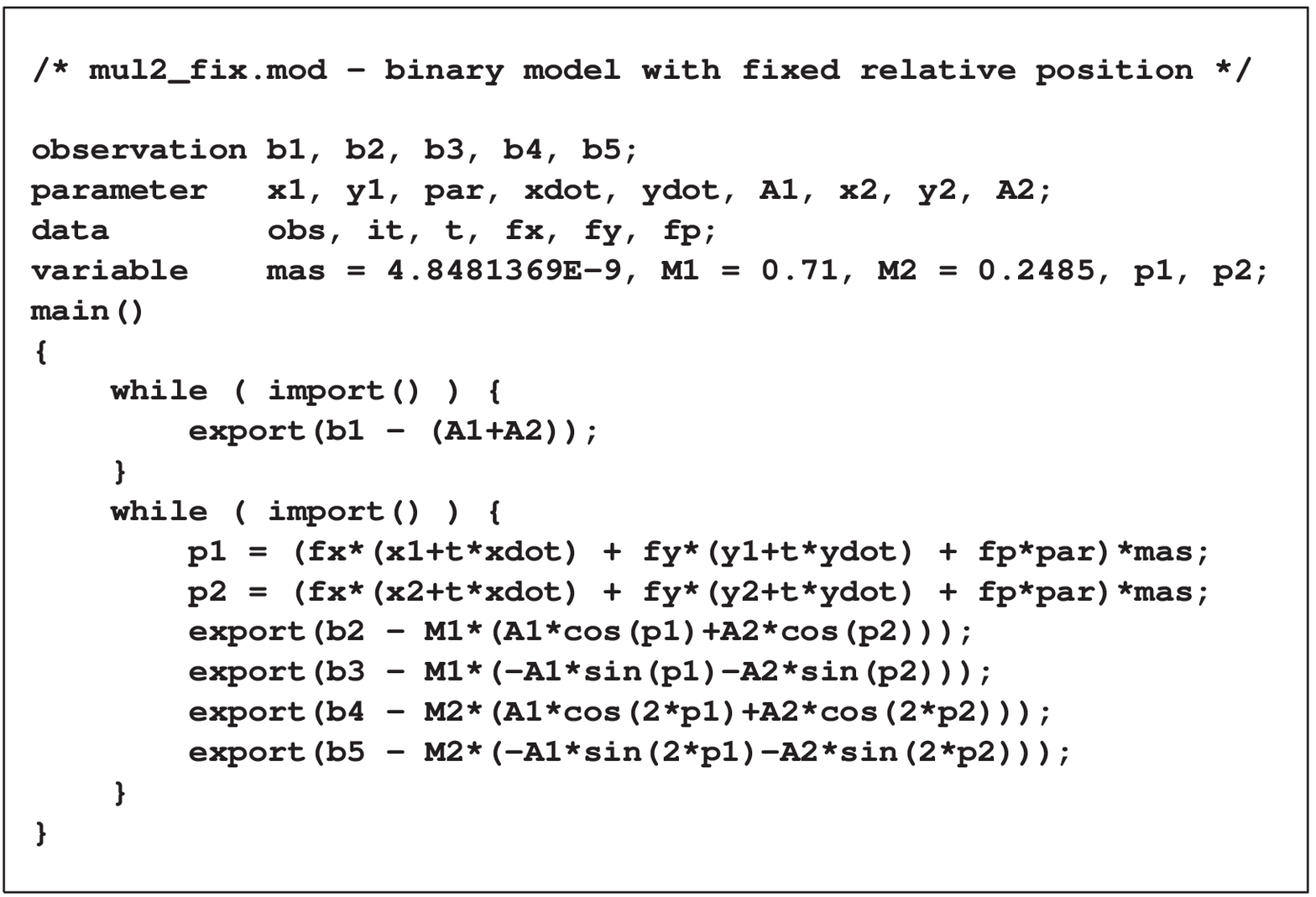}}
  \caption[ ]{An example of a double-star model defined in the 
GaussFit programming language.}
\label{fig:model}
\end{figure}

To run GaussFit, the user must supply several input files.
During execution these files are read (and sometimes modified)
by GaussFit, and additional output files generated.
For application to the TD model fitting the following input 
files are required.
\begin{itemize}
\item 
The {\it data file\/}: this contains the observational data,
in our case the TD.  A special program (td2uv.f) is available 
(Sect.~\ref{sec:soft}) to extract the TD for a given HIP number and 
format them as required by GaussFit.  The resulting data file 
consists of 16 columns and one data line per transit.  
The columns contain a sequential number for the transit, 
the target position index (JT1), the time of the transit, 
the spatial frequencies $f_x$, $f_y$, $f_p$, the signal parameters 
$b_1$--$b_5$ and their variances.  The header of the data 
file defines the name of the variable associated with each column.
\item
The {\it model file\/} (cf.\ Fig.~\ref{fig:model}): this is a
mathematical description of the object model written in the
GaussFit programming language.  This language is modeled on C, 
but includes some specific constructs.  
For instance, the declaration of variables distinguishes between
`observations' (input data with random errors that need to be 
taken into account in the fitting), `data' (error-free input data),
`parameters' (to be adjusted by the program), and ordinary
`variables'.  The special function $\tt import()$ reads one line 
of data from the data file.  The function $\tt export(x)$ sends
the equation of condition $x=0$ to the estimation algorithm, 
taking into account the uncertainties of the observational data 
that went into calculating $x$.
\item
The {\it parameter file\/}: this contains the initial guesses
of all the model parameters to be estimated.  On output it contains
the estimated parameter values and estimated errors.
\item
The {\it environment file\/} contains general information needed
for the model fitting, such as the names of the data, parameter 
and output files; the type of estimation algorithm to be used 
(standard least squares or a robust method), and stopping rules 
for the iterations.
\end{itemize}
The reader is referred to the GaussFit User's Manual (Jefferys 
et al.\ \cite{gfman}) for detailed information.

\begin{figure}[t]
  \resizebox{\hsize}{!}{\includegraphics*{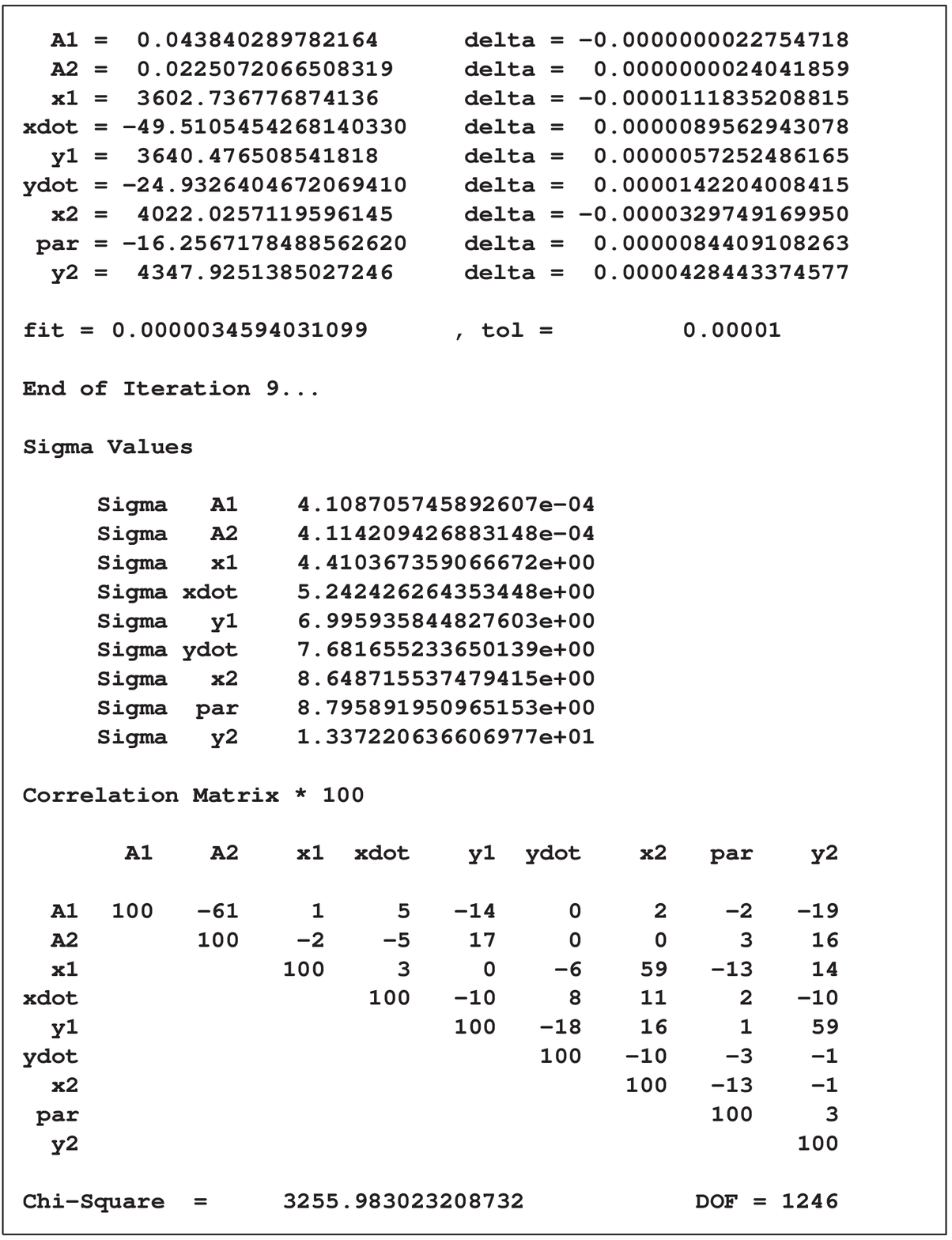}}
  \caption[ ]{Part of the GaussFit output (slightly edited) obtained 
while fitting the double star model in Fig.~\ref{fig:model} to the 
TD for HIP~97237.}
\label{fig:output}
\end{figure}

Figure~\ref{fig:model} is an example of a GaussFit model file.
It describes a binary with a fixed positional offset between the
components (i.e.\ a long-period binary).  The model parameters
are thus the astrometric parameters of the primary relative to
the reference point (${\tt x1}=\Delta\alpha*_1$, 
${\tt y1}=\Delta\delta_1$, ${\tt par}=\Delta\pi$,
${\tt xdot}=\Delta\mu_{\alpha*}$, ${\tt ydot}=\Delta\mu_\delta$),
the position of the secondary relative to the reference point
(${\tt x2}=\Delta\alpha*_2$, 
${\tt y1}=\Delta\delta_2$); and the intensities of the
components, ${\tt A1}=A_1$, ${\tt A2}=A_2$.
The components are assumed to have the same parallax and proper
motion.  The expressions within the $\tt export()$ functions
are easily recognized as the equations of condition, 
Eq.~(\ref{eq:phaseelem}), written in terms of the model parameters.
The five $\tt export()$ statements are divided among two
$\tt import()$ loops (which means that the data file is forced
to be read twice in each iteration): the reason is that GaussFit 
in its standard distribution version cannot handle more than 
four simultaneous equations of condition.

The model in Fig.~\ref{fig:model} was applied to the TD of
HIP~97237, using as starting approximation 
(3600, 3600, 0, 0, 0, 0.04, 4000, 4300, 0.02) for the variables 
in the parameter list (cf.\ Sect.~\ref{sec:images}).  The `fair'
metric with an asymptotic relative efficiency of 0.95 was 
used for robust estimation of the parameters 
(Jefferys et al.\ \cite{gf}).  Part of the output file, containing 
the results of the final (10th) iteration, is shown in
Fig.~\ref{fig:output}.  It should be noted that the estimated 
standard errors (sigma values) given in the output file have 
already been scaled by $(\chi^2/\nu)^{1/2}$, using the chi-square 
($\chi^2$) and degrees of freedom ($\nu$) given at the end of the 
file.  Adding the results of the model fitting to the reference 
point data (Sect.~\ref{sec:images}) and using the magnitude 
conversion formula $Hp=-2.5\log(A/K)$ we obtain the following
estimated parameters of the binary HIP~97237 (ICRS, epoch J1991.25):
\begin{eqnarray*}
  \alpha_1 &=& 296.43961803~\mbox{deg} 
    \pm \phantom{0}4.41~\mbox{mas}\, , \\
  \alpha_2 &=& 296.43974889~\mbox{deg} 
    \pm \phantom{0}8.65~\mbox{mas}\, , \\
  \delta_1 &=& \phantom{0}27.12836895~\mbox{deg} 
    \pm \phantom{0}7.00~\mbox{mas}\, , \\
  \delta_2 &=& \phantom{0}27.12856546~\mbox{deg} 
    \pm 13.37~\mbox{mas}\, , \\
  \pi &=& \phantom{-00}80.74 \pm 8.80~\mbox{mas} \\
  \mu_{\alpha*} &=& \phantom{00}\mbox{$-74.94$} 
    \pm 5.24~\mbox{mas~yr}^{-1}\, , \\
  \mu_\delta &=& -1203.93 \pm 7.68~\mbox{mas~yr}^{-1}\, , \\
  Hp_1 &=& 12.88 \pm 0.01~\mbox{mag}\, , \\
  Hp_2 &=& 13.60 \pm 0.02~\mbox{mag}\, . 
\end{eqnarray*}
These data are in reasonable agreement with the values derived by 
S{\"o}derhjelm (\cite{ss99}) in an orbital solution combining the TD with 
ground-based speckle observations.

\section{Software availability}
\label{sec:soft}

Fortran programs interfacing the TD with aperture synthesis software
and with GaussFit are available via the Lund Observatory Internet address 
http://www.astro.lu.se/$\sim$lennart/TD/.
The program td2uv.f extracts the TD for a given HIP identifier and 
converts them into a UV-FITS file that can be used e.g.\ by Difmap.
The program td2gf.f similarly extracts TD data and generates
a data file suitable as input for GaussFit.  Sample data files,
additional information on the TD (including descriptions of the
known errors), and links for retrieving aperture synthesis software 
and GaussFit are also given at this site.

\section{Conclusions}
\label{sec:concl}

The Hipparcos Transit Data, published as part of the Hipparcos and
Tycho Catalogues (ESA \cite{hip}), provide data from an intermediate
 step in the data
reduction process of the NDAC data analysis consortium.  Transit
Data are given for all known or suspected double or multiple star 
systems in the Hipparcos Catalogue, or about a third of the objects
in the catalogue.  There were several reasons to include these data
in the published catalogue.  A main reason was the realization that
every stellar system could not be examined for every possible type of 
solution within the time available for completing the catalogue.  
Ideally, such examination should also take into account 
ground-based observations.  The Transit Data allow
the user to re-examine such solutions, should the need arise.

Basically the Transit Data contain the Fourier coefficients which 
describe the modulation of the detector signal caused by the object's 
motion across the modulating grid.  As such they retain all the
photometric and astrometric information on the object gathered
during its transit across the grid.  The Transit Data have been
carefully calibrated and referred to the Hipparcos photometric and
astrometric systems, so that any data derived from them should be
directly comparable with other results in the published catalogue.

By giving this review of how the Transit Data were recorded, what they
physically represent and examples of their practical uses, we hope to 
encourage readers to utilize the data in 
their own exploitations of the Hipparcos results.  To aid this process, 
we have made programs available which provide interfaces with 
publicly available software packages, in particular Difmap and GaussFit.
Many other applications could be thought of -- the combination of Transit
Data with ground-based speckle observations of double stars to improve
orbits, parallaxes and mass ratios by S{\"o}derhjelm (\cite{ss99}) is an
example.  What has been covered and demonstrated here might just be a 
stepping stone to new and creative exploitations of the Transit Data.

\begin{acknowledgements}
Part of this work was supported by the Kungl.~Fysiografiska
S\"{a}llskapet i Lund and the Swedish National Space Board.  We would also 
like to thank John Conway for his expertise and help in adapting the
Transit Data for aperture synthesis imaging and to Martin Shepherd for his
help with Difmap.
\end{acknowledgements}

\end{document}